\documentclass[preprint,aps,showpacs]{revtex4}
\usepackage{epsfig}
\RequirePackage{amssymb,amsmath}
\usepackage[latin1]{inputenc}
\usepackage{graphicx}
\usepackage{indentfirst}
\usepackage{color}
\begin{document}

\title{Proton-induced fission on $^{241}$Am, $^{238}$U and $^{237}$Np 
at intermediate energies}

\author{A. Deppman}
\email{deppman@if.usp.br}
\affiliation{Instituto de Fisica, Universidade de S\~ao Paulo,
P. O. Box 66318, 05389-970 S\~ao Paulo, SP, Brazil}
  
\author{E. Andrade-II}
\email{esegundo@if.usp.br}
\affiliation{Instituto de Fisica, Universidade de S\~ao Paulo,
P. O. Box 66318, 05389-970 S\~ao Paulo, SP, Brazil}

\author{V. Guimar\~aes}
\email{valdirg@if.usp.br}
\affiliation{Instituto de Fisica, Universidade de S\~ao Paulo,
P. O. Box 66318, 05389-970 S\~ao Paulo, SP, Brazil}

\author{G. S. Karapetyan}
\email{ayvgay@ysu.am}
\affiliation{Instituto de Fisica, Universidade de S\~ao Paulo, 
P. O. Box 66318, 05389-970 S\~ao Paulo, SP, Brazil}

\author{A. R. Balabekyan}
\email{balabekyan@ysu.am}
\affiliation{Yerevan State University, Alex Manoogian 1, Yerevan 0025, Armenia}

\author{N. A. Demekhina}
\email{demekhina@nrmail.jinr.ru}
\affiliation{Yerevan Physics Institute, Alikhanyan Brothers 2, Yerevan 0036, 
Armenia\\
Joint Institute for Nuclear Research (JINR), Flerov Laboratory of Nuclear 
Reactions (LNR), Joliot-Curie 6, Dubna 141980, Moscow, Russia}

\begin{abstract} 
Intermediate energy data of proton-induced fission on $^{241}$Am, $^{238}$U 
and $^{237}$Np targets were analysed and investigated using the computational 
simulation code CRISP.  Inelastic interactions of protons on heavy nuclei 
and both symmetric and asymmetric fission are regarded. The fission 
probabilities are obtained from the CRISP code calculations by means of the 
Bohr-Wheeler model. The fission cross sections, the fissility and the number 
of nucleons evaporated by the nuclei, before and after fission, are calculated 
and compared with experimental data.  Some of the model predictions agree 
completely  with the data. We conclude that our two step model of the CRISP 
code provides a good description of intermediate energy proton-induced 
fission.
\end{abstract}
\maketitle

\section{Introduction}

The interaction between high energy protons and atomic nuclei has been 
a subject of study  over the last seventy years. Such a continuous interest 
in this subject is caused by many reasons. First of all, proton-induced 
nuclear reactions involve fundamental problems; the nucleon-nucleon 
interaction and the properties of the nuclei in various conditions of 
excitation. Modification of the proton energy and/or target nucleus 
leads to a  rich spectrum of phenomena, which have to be understood and 
described theoretically. In addition, since high energy proton collisions 
with atomic nuclei do not cause significant compression of the nuclei, 
the description of proton-induced reactions is less complex than those 
induced by heavier ions, and therefore can be useful 
in the comprehension of reactions induced by the latter probes.

The study of the proton-nucleus collision is a source of information for 
scientific and  technological applications, as for instance, Medical Physics 
applications and  Nuclear Reactor technologies. However, a broad range of 
proton energies, from a few MeV to tens of GeV and a full list of target 
nuclei must be studied. At the present, the situation, experimentally as well 
as theoretically, is rather puzzling. Despite a long history of 
investigations of proton-nucleus reactions, neither the predictive power of 
the available theoretical models provides the demanded accuracy, nor the 
experimental databases are rich enough to serve as benchmarks, which may put 
very restrictive demands and constraints on the theoretical description. 

In the case of the fission process, a comparison of calculations 
with the measured  charge, mass, energy, and spin distributions of the 
fragments, as well as the systematization of the experimental data within 
various model 
representations, can provide relevant information about the properties of 
primary fragments and the mechanism of their formation. The present work 
aims to show the results of the calculations using the CRISP code
with multimodal model in describing proton-induced fission at 660 MeV on  
heavy targets ($^{241}$Am, $^{238}$U, $^{237}$Np). The data 
considered here are from the experiments of refs. \cite{Nina3} and \cite{Nina4}.

\section{Methodology}

Presently, the only way a complete description of particle collisions, in a
large range of incident energy and on target nuclei with the mass varying 
from light nuclei as carbon to heavy nuclei as Americium, can be 
achieved,  is by considering simulation with a Monte Carlo method.
In this work we use the Monte Carlo simulation code CRISP to calculate 
the nuclear processes triggered by the inelastic interaction of protons 
with  heavy target nuclei. 
This code has been developed for more than ten years 
\cite{Deppman04,Kodama,Goncalves97,Deppman01,Deppman02a,Deppman02b,Deppman06} 
and it has been applied in the study of fission induced by photons, 
electrons and protons, and for the study of hyper-nucleus decay \cite{Israel}. 
Also, it has been used in the development of new Nuclear Reactor 
technologies \cite{Anefalos08,Anefalos05a,Mongelli}.

The main feature of this Monte Carlo code is the precise 
description of the intranuclear cascade, where a time-ordered sequence of 
collisions is governed by strict verification of the Pauli principle in a 
square-well nuclear model. In this case, pre-equilibrium 
emissions are naturally considered until the complete thermalization 
of the nucleus. After the intranuclear cascade is finished, the competition 
between evaporation and fission is described by using the Weisskopf-Ewing
model \cite{wei40}, until the nucleus is too cold to emit any other particle. 
The electroweak decay will then lead the nuclear system to its final 
ground-state, although this step of the reaction is not considered in this 
work.

In each step of the evaporation chain, the nuclear excitation energy is 
recalculated by
\begin{align}
 E_x^{(f)}=E_x^{(i)}-(B+V+\varepsilon),
\end{align}
where $E_x^{(f)}$ and $E_x^{(i)}$ are the excitation energy of the final and 
initial nucleus, respectively, $B$ is the evaporated particle separation 
energy, $V$ is its Coulomb potential, and $\varepsilon$ is the mean kinetic 
energy of the emitted particle, which is fixed at 2 MeV. 
Also, the fission channel is considered at each step with branching ratios 
given by the Bohr-Wheeler model. In the case of fission, the fragments are 
generated by following the Multimodal Random Neck Rupture Model (MM-NRM) 
\cite{Brosa90}.

Theoretically, the fission process has been successfully described by the 
Multimodal - Random Neck Rupture Model (MM-NRM)\cite{Brosa90}, 
which takes into account the collective effects of nuclear deformation 
during fission with the liquid-drop model, and includes single-particle 
effects through microscopic shell-model corrections. 
The microscopic corrections create valleys in the 
space of elongation and mass number, each valley corresponding to one 
different fission mode. The yield of a fragment, characterized 
by the fragment mass number $A$ and the atomic number $Z$, is determined 
for each mode by a Gaussian distribution. 

In the following we consider that fission can take place through three modes: 
a symmetric mode (Superlong) and two asymmetric modes (Standard I and 
Standard II).  The description of fission fragment formation allows us to 
understand the influence of the nuclear structures on the nature of fission. 
For instance, Superlong mode fragments are strongly elongated with masses 
around $A_f$/2, where $A_f$ is the mass of the fissioning nucleus 
with $A_f$=$A_H$+$A_L$, where the index $H$ and $L$ stand for the heavy and 
light fragment in a fission, respectively.  Standard I mode is characterized 
by the influence of the spherical neutron shell $N_H\sim$82 and proton 
shell $Z_H\sim$50 in the heavy fragments with masses $M_{H}\sim$132-134. 
The investigation of the influence of shell effects
and pairing correlations on the fission-fragment mass and nuclear-charge 
distributions was performed by Schmidt {\it et al.} \cite{schmidt12},
where an indication of the proton shell closure $Z=54$ effect was observed.
Standard II mode is characterized by the influence of the deformed neutron 
shell closure $N_H$=86-88 and proton shell $Z_H\sim$52 in the heavy fragments 
with masses $M_H\sim$138-140, A similar approach was recently used to study
 photon-induced fission~\cite{Evandro1,Evandro2,Evandro3,Evandro4}.

In the multimodal model, the fission cross section, as a function 
of mass number, is obtained by the sum of the three Gaussian 
functions corresponding to the three modes mentioned above \cite{Younes}:

\begin{align}
 \begin{split}
\sigma_A = &
\frac{1}{\sqrt{2\pi}}\bigg[\frac{K_{1AS}}{\sigma_{1AS}}
\exp\left({-\frac{(A-A_S-D_{1AS})^2}{2\sigma^2_{1AS}}}\right)+
\frac{K'_{1AS}}{\sigma'_{1AS}}\exp\left(-\frac{(A-A_S+D_{1AS})^2}
{2\sigma'^2_{1AS}}\right)+\\
 & \frac{K_{2AS}}{\sigma_{2AS}}\exp\left({-\frac{(A-A_S-D_{2AS})^2}
{2\sigma^2_{2AS}}}\right)+
\frac{K'_{2AS}}{\sigma'_{2AS}}\exp\left({-\frac{(A-A_S+D_{2AS})^2}
{2\sigma'^2_{2AS}}}\right)+\\
 & \frac{K_S}{\sigma_S}\exp\left({-\frac{(A-A_S)^2}
{2\sigma^2_S}}\right)\bigg],
\label{mass}
 \end{split}
\end{align}
\noindent where $A_S$ is the mean mass number determining the center 
of Gaussian functions; and $K_i$, $\sigma_i$, and D$_i$ are the contribution,  
dispersion and position parameters of the $i^{th}$ Gaussian functions.  
The indexes $AS$, $S$ designate the asymmetric and symmetric components. 

The CRISP code works on an event-by-event basis, and therefore the parameter 
$A_S$ in Eq. (1) is completely determined by the mass of the fissioning nucleus
$A_f$, that is, $A_S=A_f/2$. The quantities  $A_S$ + D$_{iAS}$ = A$_H$ and 
$A_S$ - D$_{iAS}$ = A$_L$, where $A_H$ and $A_L$ are the masses
of the heavy and light fragment, respectively, determine the positions of 
the heavy and light peaks of the asymmetric components on the mass scale. 
The value of $A_S$=(A$_H$ + A$_L$)/2 is treated as the mass of the nuclei that 
undergo fission in the respective channel. 

One important observable of the fission process is the charge distribution of
a given isobaric chain with mass number $A$.  It is assumed that this
fission fragment charge  distribution is well described by a Gaussian 
function characterized by the most probable charge, $Z_p$ of an isobaric 
chain $A$ (centroid of the Gaussian function) and the associated width 
parameter, $\Gamma_z$ of the distribution as following \cite{kudo,Duijvestijn}:

\begin{eqnarray}
\sigma_{A,Z}=\frac{\sigma_A}{\Gamma_z\pi^{1/2}}
\exp\left({-\frac{(Z-Z_p)^2}{\Gamma_z^2}}\right),
\label{charge}
\end{eqnarray}

\noindent where $\sigma_{A,Z}$ is the independent cross section of the 
nuclide $Z,A$. The values $\sigma_A$ correspond to the total fission cross 
section of a given isobaric chain with mass number $A$. 
The values $Z_p$ and $\Gamma_z$ can be represented as slowly varying 
linear functions of the mass number of the fission fragments:

\begin{eqnarray}
Z_p=\mu_1+\mu_2A\,,
\label{zp}
\end{eqnarray}
\noindent and
\begin{eqnarray}
\Gamma_z=\gamma_1+\gamma_2 A\,.
\label{width}
\end{eqnarray}

The values for these parameters obtained by a fitting procedure from 
Ref.~\cite{Nina3,Nina4} are: $\mu_1=4.1$, $\mu_2=0.38$, $\gamma_1 = 0.92$ 
and $\gamma_2=0.003$ for $^{241}$Am and $\mu_1=5.0$, $\mu_2=0.37$,  
$\gamma_1 = 0.59$ and $\gamma_2=0.005$ for $^{237}$Np. In the present work, 
we applied the same values obtained for $^{241}$Am to $^{238}$U.

Analysis using Eqs. (\ref{mass}), (\ref{charge}), (\ref{zp}) and (\ref{width}) 
has been performed with success to describe fission induced by different 
probes; thermal-neutrons \cite{Hambsch02,Hambsch03}, protons up to energies 
of 190 MeV \cite{Duijvestijn,ohtsuki}, 200 MeV neutrons \cite{Maslov2003},  
and heavy-ions \cite{Itkis,Pokrovsky}. In these works, the yield,  position, 
and width parameters for each mode  in Eq.  (\ref{mass})  were considered as 
free parameters in the fitting procedure. Here we use the multimodal model 
associated with the Monte Carlo code CRISP, which simulates the entire 
process up to the point of fission. In the CRISP code, the fissioning 
nucleus of all events is known and, therefore,  the mass of the perfectly 
symmetric fission fragments is given by $A_S$=$A_f/2$. 

Whenever the fission channel is chosen, the masses and atomic numbers of the 
heavy fragments produced, $A_H$ and $Z_H$, respectively, are sorted according 
to Eq. (\ref{charge}).  The light fragments are obtained according to 
$A_L=A_f-A_H$ and $Z_L=Z_f-Z_H$, where $Z_f$ is the atomic number of the 
fissioning system.

As a final step, all fragments obtained go into a final
 evaporation step according to the model of evaporation/fission competition 
already mentioned. The energy of each fragment is determined using:
\begin{align}
E_i = \dfrac{A_i}{A_f} E_{frag},
\label{energy}
\end{align}

\noindent where $E_{i}$ and $A_{i}$ are the excitation energy and the mass 
number of the fragment $i$, respectively. $E_{frag}$ is the total excitation 
energy of the fragments, which is assumed to be equal to the excitation 
energy of the fissioning system.

Recently, this method of simulating fission reactions was used in the 
analysis of photo-fission with bremsstrahlung photons at end-point energies 
50 MeV and 3500 MeV on $^{238}$U and $^{232}$Th targets, with satisfactory 
results~\cite{Deppman13}.

\section{Results and Discussion}

\subsection{Mass Distribution}

The results for fragment mass distributions obtained with the CRISP  code are 
presented in Fig. 1. Results of the best fitted distributions from 
Ref. \cite{Nina3, Nina4} are also shown in the figure for comparison. 
The calculated results from the CRISP code are obtained with the parameters 
shown in Table \ref{tabParam}. As can be observed, both fit and calculated 
distributions reproduce the shape of the experimental distributions. 
The calculated position and width of the peaks for symmetric and asymmetric 
modes are in fair agreement with the experimental distributions. However, the 
calculated distributions by the CRISP code  are systematically below the 
data 
for the $^{241}$Am and $^{237}$Np targets,  indicating that the calculations 
for total fission cross sections underestimate the experimental cross sections. 

The experimental total fission  cross section is estimated by:
\begin{align}
 \sigma_F^{exp} = \dfrac{1}{2} \sum_i \sigma^{exp}(A_i),
\end{align}
where $\sigma^{exp}(A_i)$ is the experimental cross section for each mass 
number $A$, where the factor $\dfrac{1}{2}$ has to be considered to avoid 
double counting of fission events due to the summation over the fragments. 

The CRISP code calculates the total fission cross section by supposing 
that it is given by $\sigma_{F}^{calc} = D\sigma_{in}$,  where $D$ is the 
nuclear fissility and $\sigma_{in}$ is the total cross section for the 
inelastic interaction. The CRISP code adopts the geometrical cross section 
to estimate the inelastic cross sections. 
\begin{align}
\sigma_{in} \sim \sigma_{g} = \pi \left( r_0 + r_0 A^{1/3} \right)^2\,,
 \label{geomCS}
\end{align}
In the geometric cross section, the nucleus is considered as a sphere 
with radius $R(A) = r_0 A^{1/3}$ and the proton as a sphere with 
radius $r_0$, where $r_0 = 1.2$ fm. The values for both experimental and 
calculated total fission cross sections obtained are shown in Table 
\ref{tabParam2}. The ratios between calculated and experimental total fission 
cross sections  $\sigma_F^{calc}/\sigma_F^{exp}$ are $0.6 \pm 0.1$, $0.8\pm 0.1$ 
and $0.7\pm 0.1$, respectively for $^{241}$Am, $^{238}$U and $^{237}$Np.

The calculated and experimental fissility, determined as the ratio 
$D = \sigma_F/\sigma_{in}$, for proton-induced fission on $^{241}$Am,  
$^{238}$U and $^{237}$Np targets as a function of the parameter $Z^2/A$ are 
plotted in Fig. \ref{figFissility} together with estimated experimental values 
for protons \cite{Nina3, Nina4} and photons \cite{Chung,Rubchenya} on 
$^{238}$U and $^{232}$Th targets. As can be noticed, the fissilities for 
fission induced by protons and photons on different targets and at different 
incident energies are below unity, and show a plateau of saturation 
for incident energies above $\sim 40$ MeV \cite{Nina1,Nina2,Fukahori}. 
We can also observe in this figure that calculated fissilities with the CRISP 
code are close to unity and above the experimental values. A possible 
explanation for this behaviour is the fact that the total inelastic 
proton-nucleus cross section is being underestimated.  The quantitative 
difference can be attributed to the geometrical approximation given by  
Eq. (\ref{geomCS}), which assumes that the nuclei are spherical, an hypothesis 
which might not hold for heavier nuclei, such as those studied here. 

\subsection{Symmetric and Asymmetric Modes}

One striking feature of the fragment mass distributions present in Fig. 1 is 
that the asymmetric fission contribution is much more evident for $^{238}$U 
than for the other nuclei studied, despite  the fact that they have similar 
masses. This behavior can be explained by taking into account the empirical
expression for the  critical value of the fissility 
parameter defined by Chung {\it et al.}  \cite{Chung}:

\begin{eqnarray}
(Z^2/A)_{cr}=35.5+0.4(Z_f-90).
\label{crit}
\end{eqnarray}
where $Z_f$ is the atomic number of the fissioning nucleus. According to 
Chung {\it et al.} \cite{Chung}, for nuclei with values of $Z^2/A$ greater than
the critical value, the symmetric fission mode is dominant, while for nuclei 
with smaller values the main fission channel leads to asymmetric fragment 
distribution. The higher the fissility parameter, with respect to the critical 
value, the higher is the  probability to obtain a symmetric mass distribution. 

The critical fissility parameter $(Z^2/A)_{cr}$ for $^{241}$Am, $^{238}$U and 
$^{237}$Np are 37.5, 36.3, and 36.7, respectively, while the average 
fissility parameter, $Z^2/A$, is 39.7 ($A=227,Z=95$) for $^{241}$Am, 
37.3 ($A=227,Z=92$) for $^{238}$U and 38.7 ($A=223,Z=93$) 
for $^{237}$Np. 

The smallest difference between $Z^2/A$ and  $(Z^2/A)_{cr}$ is found
for  $^{238}$U and it could explain the larger contribution of asymmetric 
fission in the mass distribution for this nucleus. A lower $Z^2/A$ value for 
$^{238}$U is a consequence of the pre- and post-equilibrium emissions, which 
result in compound and fissioning nucleus mass distributions rather different 
from those obtained for the other target nuclei. We present in Table 
\ref{tabParam2} the average mass of the compound nucleus $A_{CN}$, 
the average fissioning nucleus mass $A_f$ and the average mass of fission 
fragments after evaporation,$A_{ff}$, for the three cases studied here. 
The comparison between them shows that the number of pre-scission neutrons 
is higher for $^{238}$U, which could be related to a lower excitation 
energy of the fissioning nucleus. This is confirmed by the lower number of 
post-scission neutrons for uranium when compared to the other nuclei. 
Thus, although the total number of neutrons emitted is approximately the 
same for all nuclei studied here, the pre-scission evaporation chain is 
longer for $^{338}$U. This causes not only a lower excitation energy, but also 
the formation of a lighter fissioning system, with a lower $Z^2/A$ parameter, 
explaining the more pronounced contribution of asymmetric fission for $^{238}$U 
as compared to those for $^{241}$Am and $^{237}$Np.

\subsection{Proton and Neutron Emissions}

Besides the fragment mass distribution, it is also interesting to analyze 
some aspects of the fission process related to charge distributions and 
particle emission of the fragments. The charge  distribution for an isobar 
chain with mass number $A$, from a fissioning heavy nuclei, is characterized 
by a Gaussian shape given by Eq.~(\ref{charge}), with parameters,  $Z_p$ and 
$\Gamma_z$, where  $Z_p$ and $\Gamma_z$ are the most-probable charge and the 
corresponding width of the distribution. In Fig. \ref{figZp} we show the 
comparison between experimental and calculated values with the CRISP code 
for $Z_p$ as a function of $A$.  As can be seen, the calculation reproduces 
quite well the experimental data and both of them show a clear linear  
dependence of $Z_p$ with $A$, as expected by Eq. (4) and (5) \cite{kudo}. In 
Fig. \ref{figDevCharge} we plot the difference between the calculated width  
$\Gamma_p$ and  experimental values.  We observe that for $^{241}$Am and 
$^{238}$U, the data fluctuates around zero,  as one would expect, with 
standard deviations of $3.4$ and $2.6$, respectively. In the case of 
$^{237}$Np,  although the values are close to zero, the results show a clear 
linear dependence of the parameter $\Gamma_p$ with A.  In general, the 
calculated values are in good agreement with the experiment, especially 
for the fragments in the low and medium mass regions.  In the case of 
$^{237}$Np, the calculations systematically overestimate the width of the 
isotopic distributions in the range of medium and high masses.

With the CRISP code we can also obtain the average number of  pre- and 
post-scission emitted neutrons,  which are reported in Table \ref{tabParam2}.
The sum of these two contributions gives the average number of emitted 
neutrons, which can be compared with the experimental values, also shown in 
Table \ref{tabParam2}. We observe good agreement between calculated and 
experimental values, showing that the theoretical 
predictions for the emission of neutrons are correct.

The analysis of neutron emissions and atomic number distributions shows 
that the CRISP code gives a good description of the mechanisms for 
emissions of nucleons in the pre- and post-equilibrium stages of the 
nuclear  reaction. It is important to note that the number of emitted 
nucleons is directly related to the excitation energy of the compound 
nucleus formed in the reaction, therefore, we can conclude that the 
excitation energies, as calculated by CRISP, are also supported by the 
experimental results. The present analysis also indicates that our theoretical 
model gives a good description of the dynamical process taking place inside 
the nucleus during reactions at intermediate energies.

\section{Conclusion}

In this work, an analysis of the fragment mass distributions 
obtained in the fission of $^{241}$Am, $^{238}$U and $^{237}$Np induced 
by 660 MeV protons is presented. The analysis is performed by comparing the 
results from a Monte Carlo calculation with the CRISP code
with experimental data from Ref. \cite{Nina3, Nina4}.
We show that the CRISP code can give a reliable description of the fission 
dynamics for the reactions studied here. In fact, the mass distributions 
for fission fragments are correctly described by considering three fission 
modes, one symmetric and two asymmetric, for all three targets studied. 
The evaporation of fission fragments is also considered, and 
we found that this mechanism is relevant for the description of the final 
fragment masses. The pre- and post-scission neutron emission and the atomic 
number distributions were also analyzed, and we show that calculations 
and experiments are in good agreement. The information of pre- and 
post-scission neutron emissions is important in explaining the different 
relative contribution of asymmetric fission with respect to symmetric 
fission for uranium when compared to the other two target nuclei.

\section*{Acknowledgment} 

G. Karapetyan is grateful to Funda\c c\~ao de Amparo \`a Pesquisa do Estado 
de S\~ao Paulo (FAPESP) 2011/00314-0, and to International Centre for 
Theoretical Physics (ICTP) under the Associate Grant Scheme. We thank 
prof. Wayne Seale for reviewing the text.

\newpage

\begin{table}[!ht]\centering
\caption{Parameters for the mass distribution calculations.}
\label{tabParam}
\scalebox{1.0}{
\begin{tabular}{|c|c|c|c|} \hline
Parameter       & $^{241}$Am & $^{238}$U  &  $^{237}$Np	\\ 
\hline
$K_{1AS}$        &    45.0   &   53.80   &    49.0	\\
$K'_{1AS}$       &    45.8   &   52.00   &    49.0	\\ 
$\sigma_{1AS}$	&     4.2   &    1.60   &     4.5	\\
$\sigma'_{1AS}$	&     4.2   &	 1.71	&     4.5	\\
$D_{1AS}$	&    20.0   &	22.50	&    21.3	\\
$K_{2AS}$	&   220.5   &  477.32	&   252.0	\\ 
$K'_{2AS}$	&   220.5   &  476.52	&   252.0	\\
$\sigma_{2AS}$	&     7.0   &	 4.29	&     6.5	\\
$\sigma'_{2AS}$	&     7.0   &	 4.19	&     6.5	\\
$D_{2AS}$	&     25.5  &	22.90	&    26.3	\\
$K_S$		&   2970.0  & 1396.45	&  2590.0	\\
$\sigma_{S}$	&     15.0  &	14.2	&    13.7	\\ 
\hline
\end{tabular} }
\end{table}

\begin{table}[ht]\centering
\caption{Calculated and experimental quantities.}
\label{tabParam2}
\begin{tabular}{|c|c|c|c|}  \hline
Parameter                  & $^{241}$Am     & $^{238}$U      & $^{237}$Np  \\ 
\hline 
$(\sigma_{tot})_{exp}$(barn) & 1.76$\pm$0.30 & 1.23$\pm$0.18 & 1.60$\pm$0.24\\
$(\sigma_{tot})_{cal}$(barn) & 1.08          & 0.95          & 1.07 \\
$(A_S$)$_{exp}$             & 113.5$\pm$0.6 & 113.5$\pm$0.6 & 111.7$\pm$0.9\\
$(A_S$)$_{cal}$             & 113.0         & 114.0         & 111.5        \\
$(A_f$)$_{exp}$             & 227.0         & 227.0         & 223.4    \\
$(A_{ff})_{cal}$             & 226.0         & 228.0         & 223.0  \\
$(A_f)_{cal}$               & 237.7         & 232.5         &  234.7  \\
$(A_{comp})_{cal}$           & 238.9         & 233.0         & 234.9\\
(pre-scission neutrons)$_{cal}$  & 4.3      & 6.5           & 3.3 \\
(post-scission neutrons)$_{cal}$ & 11.7     & 4.5           & 11.7 \\
(evaporated neutrons)$_{exp}$    & 15$\pm2$ & 12$\pm$2      & 15$\pm$2 \\ 
\hline
\end{tabular} 
\end{table}

\begin{figure}
\epsfig{file=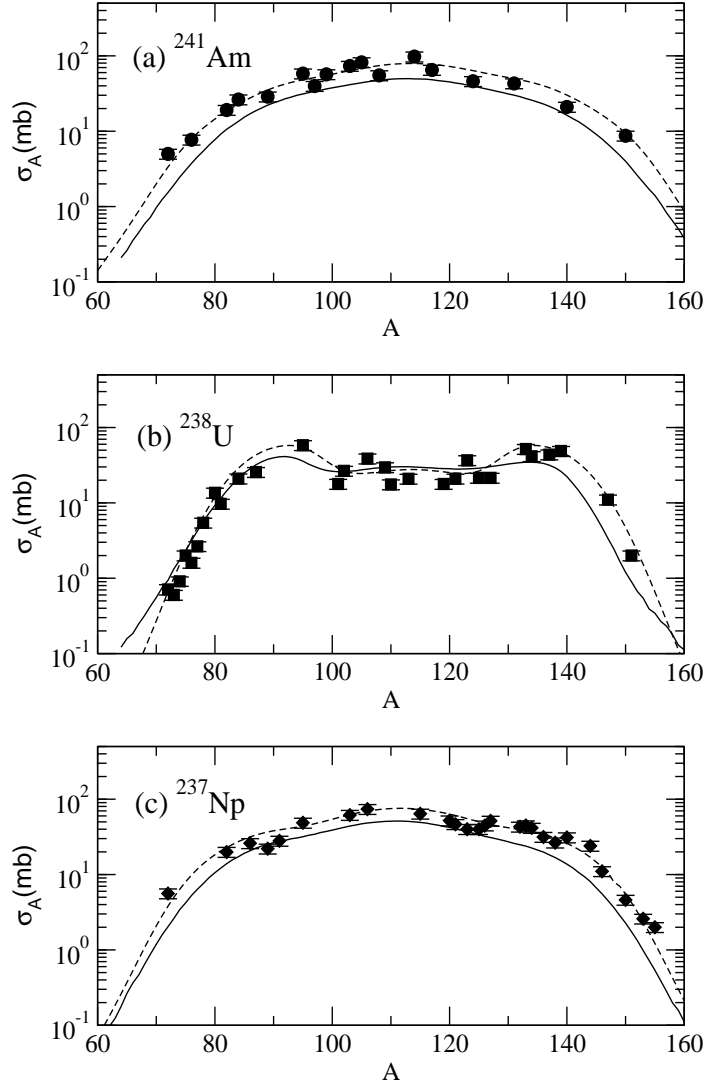,height=16cm,width=12cm,angle=0.}
\caption{Mass distributions of fission fragments induced by protons at 
$E_{p}=660$ MeV on a) $^{241}$Am, b) $^{238}$U  and c) $^{237}$Np targets.
The solid symbols; square, circle and triangle are the experimental cross 
sections in units of  milibarn ($mb$) for each of the indicated targets
as a function of the isobaric chain $A$. The solid black line corresponds to 
the calculation by the CRISP code and the dotted line gives the results of a 
minimum $\chi$-square fitting over the experimental data from Refs. 
\cite{Nina3} and \cite{Nina4}.}
\label{figMassDist}
\end{figure}

\begin{figure}
\epsfig{file=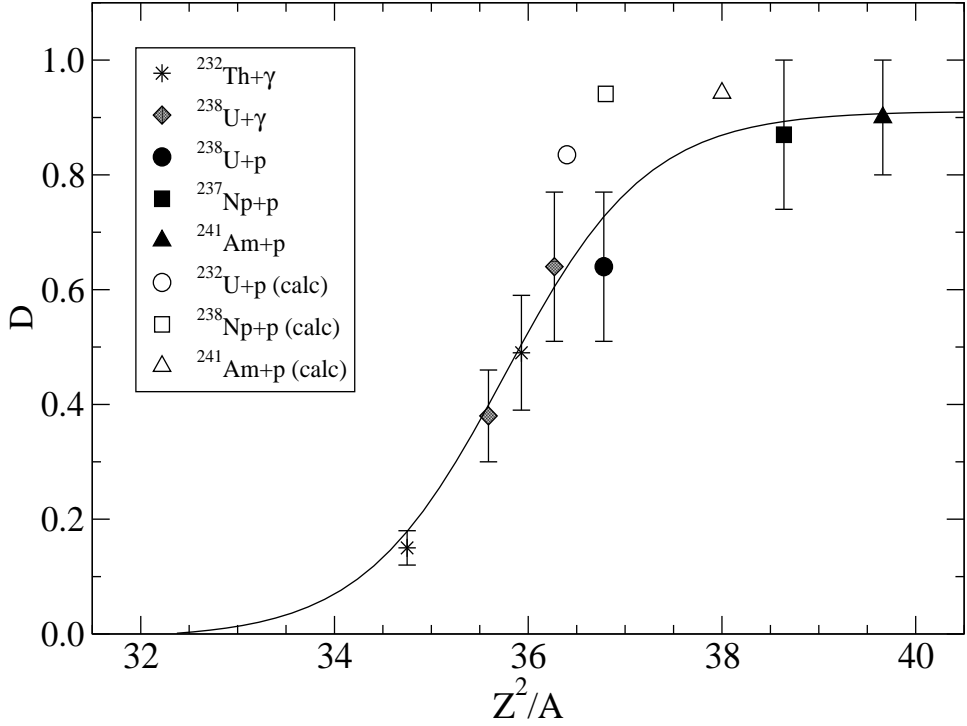,height=15cm,width=12cm,angle=-90.}
\caption{Fissility D as a function of $Z^2$/A for p+$^{237}$Np, p+$^{238}$U and  
p+$^{241}$Am (calculated from the present work and experimental data from 
Refs. \cite{Nina3,Nina4}), $\gamma$+$^{238}$U and $\gamma$+$^{232}$Th  from 
Refs. \cite{Chung,Rubchenya}, as indicated.  Calculations by CRISP are open 
symbols without error bars.  The solid line is to guide the eye to the 
experimental points.}
\label{figFissility}
\end{figure}

\begin{figure}
\epsfig{file=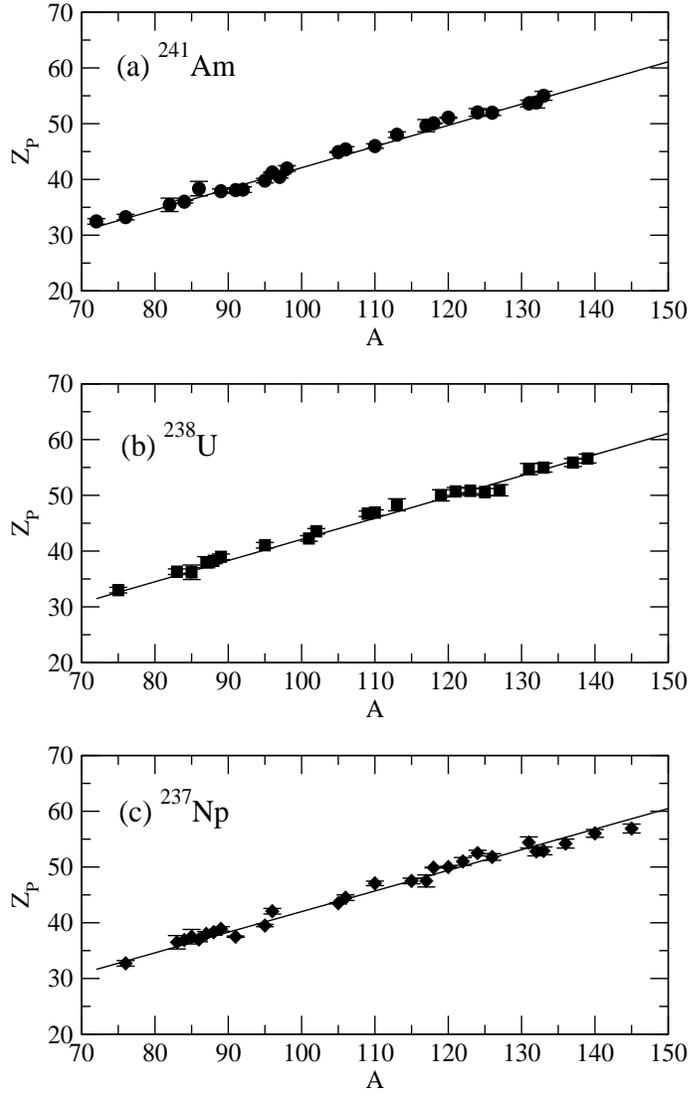,height=16cm,width=12cm,angle=0.}
\caption{The most probable charge $Z_{p}$ for a) $^{241}$Am, 
b) $^{238}$U and c) $^{237}$Np  targets, respectively. The calculations 
by CRISP are given by the solid line while the filled  symbols 
are experimental data.}
\label{figZp}
\end{figure}

\begin{figure}
\epsfig{file=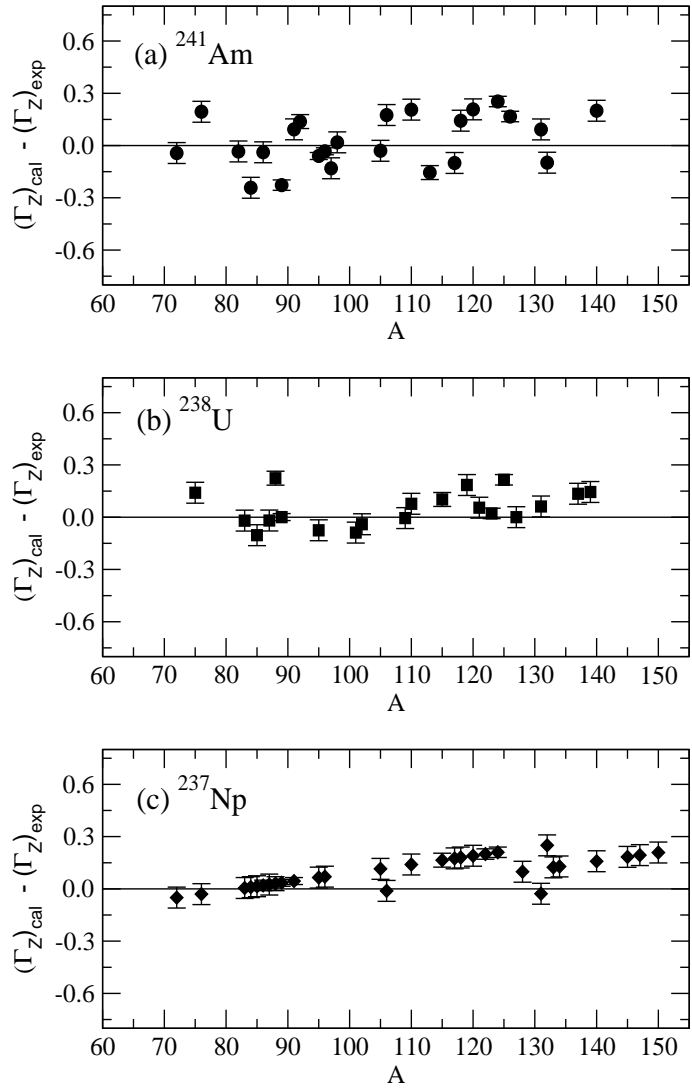,height=16cm,width=12cm,angle=0.}
\caption{The deviation between the  experimental width of charge 
distribution  and the values calculated by the CRISP code for a) $^{241}$Am, 
b) $^{238}$U and c) $^{237}$Np targets.}
\label{figDevCharge}
\end{figure}



\end{document}